\documentclass[prd,aps,showpacs,groupedaddress,eqsecnum,notitlepage,nofootinbib,twocolumn]{revtex4-2}
\usepackage{graphicx}
\usepackage{tikz,siunitx,mwe}
\usepackage{tensor}
\usepackage{epstopdf}
\usepackage{amsmath}
\usepackage{amsfonts}
\usepackage{slashed}
\usepackage{amssymb}
\usepackage{enumerate, color}
\usepackage{subfigure}
\usepackage{lipsum}
\usepackage{color}
\usepackage{graphicx,bm}
\usepackage[utf8]{inputenc}
\usepackage{eurosym}
\usepackage{scalerel}
\usepackage{float}
\usepackage{accents}
\usepackage[colorlinks=true,pdfstartview=FitV,linkcolor=blue,citecolor=blue,urlcolor=blue,breaklinks=true]{hyperref}
\newcommand{\be}{\begin{equation}}
\newcommand{\ee}{\end{equation}}
\newcommand{\ben}{\begin{eqnarray}}
\newcommand{\een}{\end{eqnarray}}
\newcommand{\bes}{\begin{subequations}}
\newcommand{\ees}{\end{subequations}}

\newcommand{\sech}{{\rm sech}}

\begin{document}
\title{Analytical scalar field solutions on Lifshitz spacetimes}
\author{Danilo C. Moreira}\email{moreira.dancesar@gmail.com}
\affiliation{Unidade Acad\^emica de F\'isica, Universidde Federal da Campina Grande, 58109-970, Campina Grande-PB, Brazil,}
\affiliation{N\'ucleo de Forma\c c\~ao de Docentes, Universidade Federal de Pernambuco, 55014-900, Caruaru, PE, Brazil}
\begin{abstract}
In this work, we investigate the existence of analytic solutions of static scalar fields on Lifshitz spacetimes. We evade Derrick's theorem on curved spacetimes by breaking general covariance and use first-order formalism to obtain solutions with finite energy related to the time-translational invariance of the background geometry along with the energy-momentum tensor of the model. We show that such solutions exist and are stable in systems where the Lifshitz background geometry is fixed and the self-interaction potential of the scalar field explicitly depends on the radial coordinate present in the metric.
\end{abstract}

\maketitle


\section{Introduction}

Obtaining and analyzing soliton and solitonlike topological structures in field theory has been a topic discussed for some time in the literature and constitutes a branch of physics with many applications, where it is possible to infer ideas about several interesting problems \cite{vilenkin2000cosmic,manton2004topological,vachaspati2006kinks}. An important point about the stability of these solutions is the Derrick's scaling theorem \cite{derrick1964comments,hobart1963instability}, which under certain conditions asserts the impossibility of the existence of nonzero finite-energy static solutions in scalar field theories in flat spaces with more than one spatial dimension. In particular, in 1+1 dimensions one can find nontrivial topological solutions as kinks in scalar field models if the scalar interaction potential is  non-negative and holds some nonunitary set of degenerate minima, which forms topological sectors where kink-antikink pairs are located - a pedagogical review on this subject is found in \cite{bazeia2005defect}.

Derrick's theorem can be evaded in some ways by breaking some of its requirements and an interesting way to do this is presented in \cite{bazeia2003new}, where is proposed a setup with classical scalar fields living in a D-dimensional flat spacetime with a scalar interaction potential explicitly depending on the background coordinates in a nontrivial way. In this context, the set of relativistic symmetries is broken due to the violation of the momentum invariance which comes from the emergence of a preferential point in space arising from the choice made for the scalar potential. Furthermore, in that setup is also possible to build a first-order formalism which gives rise to stable BPS solutions and also acquire conserved charges related to the topological profile of the model. For the convenience of presenting a systematic way to achieve topological solutions by using first-order equations in planar systems with radial symmetry, the formalism presented in \cite{bazeia2003new} was generalized \cite{bazeia2005global,casana2015trapping} and recently has been very useful in modeling analytical solutions of vortex \cite{bazeia2018maxwell,bazeia2018stable,casana2020bps,andrade2019first,bazeia2019multilayered} and  monopoles \cite{bazeia2018magnetic,bazeia2021novel} equipped with magnetic permeability and internal structure, in addition to Skyrmion-like structures \cite{bazeia2016topological,bazeia2017topological,bazeia2017semi,bazeia2019configurational,bazeia2021configurational}. It has also led to interesting results regarding fermionic spectra in boson-fermion systems \cite{casana2014trapping,bazeia2018dirac}.

Another way to evade Derrick's theorem is by setting up systems where a classical scalar field is inserted into a curved background with nonbackreacting geometry. This way of looking for solitonic solutions is different from those covered by Derrick's theorem and was first addressed in the late 1970s, when this idea was applied to models with scalar fields on charged black holes \cite{radmore1978non,palmer1979derrick}. However, in these systems nontrivial solutions may still exist within a certain small range of mass and charges if one relaxes the conditions imposed on the scalar potential. Recently, this subject has been taken up and new studies about the existence and properties of static scalar field solutions on radially-symmetric curved spacetimes have been used to study the interaction between global defects and black holes \cite{perivolaropoulos2018gravitational,alestas2019evading}.

A proposal to generalize Derrick's theorem for static asymptotically flat spacetimes with no backreaction was presented in \cite{carloni2019derrick}, where the authors also extend the result by adding backreaction from some scalar-tensor models. However, a way to violate it using first-order equations and finding BPS solutions has already been presented in \cite{morris2021radially} through a generalization of the approach presented in \cite{bazeia2003new} to static and asymptotically flat spacetimes with radial symmetry. 
It is also possible to evade the Derrick's theorem generalization proposal in \cite{carloni2019derrick} in compact Einstein spacetimes \cite{hartmann2020real}, which can represent a FRW universe for small time scales. In this case, the compact structure of the metric arising from the imposition of a length scale on the model induces the necessary scenario for the formation of solitonic solutions with real scalar fields. This idea of inducing geometric constriction effects to study the formation of kinklike solutions also appears in recent studies in 1+1 dimensions (but in models covered by Derrick's theorem) with applications in the analysis of bound states of fermionic spectra of models derived from the Jackiw-Rebbi model \cite{jackiw1976solitons,bazeia2020geometrically,bazeia2021fermions,bazeia2019fermion,bazeia2017fermionic}.
 
 In all the cases mentioned above, Derrick's theorem is violated by imposing that the static scalar field is situated on a background geometry with axial symmetry and no backreaction. Furthermore, the models mentioned so far also have in common the fact that they use real scalar fields in asymptotically flat background geometries presenting isotropic scaling in coordinates. In the present work we study the formation of static and spatially localized scalar field solutions on Lifshitz spacetimes, which arise in the context of the gauge/gravity duality \cite{maldacena1999large,ammon2015gauge}
 as a class of spacetimes such that their dual field theories are nonrelativistic and present anisotropic scaling of the Lifshitz type \cite{kachru2008gravity,taylor2016lifshitz}. In this case, the gravity side of the Duality presents a nonasymptotically flat geometry derived from Einstein equations in the presence of massive vector fields which has anisotropic scaling symmetry. Hence, it is not covered by Derrick's theorems and the extensions formulated to date.  Due to its scaling properties, Lifshitz spacetimes have been explored in different contexts, constituting a great source of applicability (see, for instance, models involving thermodynamics \cite{brenna2015mass,balasubramanian2009analytic,ayon2010analytic,ayon2009lifshitz,bertoldi2009black,brito2020black,natsuume2018holographic,bazeia2015two,deveciouglu2011thermodynamics} and microscopic entropy counting \cite{gonzalez2011field,melnikov2019lifshitz,bravo2020thermodynamics,ayon2019microscopic} in asymptotically Lifshitz solutions and other related references). In gauge/gravity duality, the mass of the scalar field in bulk is associated to the scaling dimension of the scalar operators in the boundary and by dealing with non-relativistic setups we can provide ways to approach nonrelativistic field theories with applications in condensed matter systems \cite{hartnoll2009lectures}. The action we use describes a standard classical scalar field, so the related dual operator cannot provide values with short range (UV) corrections arising from possible contributions derived from Lifshitz-type anisotropies on the scalar sources, but only from those anisotropies associated with the background geometry. Moreover, the usual kinetic term in the scalar action is relevant because it provides second-order equations and the required structure for the energy-momentum tensor so that the BPS formalism can be settled. Therefore, extending studies on this subject can still lead to a wide range of applications and a better understanding of the behavior of scalar fields in this context can bring relevant ingredients to the development of new results.

This work is organized as follows. In the next section we discuss properties of Lifshitz spacetimes, present the action of the model and discuss the basic properties of the scalar potential and field equations along with its boundary conditions. In Sec. III we present the BPS formalism used to reduce the order of the equations of motion and find solutions and associated conserved charges. In Sec. IV we provide an illustrative example of obtaining analytic solutions in Lifshitz spaces and, finally, in Sec. V we show that the models coming from these systems are stable. We end with some final comments about the results obtained.

\section{Static scalar fields on Lifshitz spacetimes}

Lifshitz spaces were introduced in \cite{kachru2008gravity} and have its $D$-dimensional metric given by
\begin{equation}\label{lifmetric}
ds^2=-\left(\frac{r}{\ell}\right)^{2z} dt^2+\left(\frac{\ell}{r}\right)^2 dr^2+\left(\frac{r}{\ell}\right)^2 dx^i dx^i,
\end{equation}
where $(x^0,x^1)\equiv (t,r)$, the $r$-coordinate is a radial one, $\ell$ is a length scale, $z$ is a parameter know as the {\it dynamical exponent}, which represents a measure of the anisotropy presented by the system, and $i=2, 3, \cdots, D$. It satisfies the following set of nonrelativistic symmetries:
\bes\label{nonrelsym}\ben
\label{timeinv} H:t&\rightarrow& t'=t+a,\\
P^{i}:x^{i}&\rightarrow&x'^{i}=x^{i}+a^{i},\\
L^{ij}:x^{i}&\rightarrow&x'^{i}=L^{i}_{~j}x^{j},
\een\ees
which represent invariances under temporal and spatial translations as well as spatial rotations with $L^{i}_{~j}\in SO(D-2)$, respectively
\cite{taylor2016lifshitz}. In the particular case where $z = 1$ we retrieve anti-the-Sitter ($\text{AdS}_{D}$) spacetime in Poincar\'e coordinates and for $z=2$ the symmetries \eqref{nonrelsym} belong to the Schr\"{o}dinger group. For $ z \neq1 $ Lifshitz spacetimes do not appear as gravitational solutions in scalar-tensor theories, since they are only obtained in the presence of vector fields. The metric \eqref{lifmetric} is also invariant under the anisotropic scaling
\begin{equation}\label{lifscal}
\mathcal{D}_{z}: t\rightarrow \beta^z t, ~x^i\rightarrow \beta x^i~~\text{and} ~~ r\rightarrow r/\beta,
\end{equation}
which gives the radial coordinate a key role for applications, since it is adjusted in such a way that, in holographic models, the resulting geometry in the gravity side is dual to quantum systems with Lifshitz invariance and energy scale set by $r$. In the original models where the Lifshitz geometries emerged, the condition $z\geq1$ is necessary for well-behaved massive vector fields used to generates these solutions. However, this background also arises in models as Ho\v{r}ava-Lifshitz gravity \cite{griffin2013lifshitz}, where the values the $z$-parameter can assume are arbitrary. Therefore, as we are dealing here only with the scaling properties of the geometry itself, we assume that the dynamical exponent lies on the line $ - \infty <z <\infty $.

In this work we are interested in studying properties of classical scalar fields on Lifshitz spacetimes. We adopt the strategy of analyzing scalar fields defined by an effective action where the potential depends on the background geometry coordinates, given by 
\begin{equation}\label{action}
S_{\left(\phi\right)}=\int d^D x\sqrt{-g}\left(-\frac{1}{2}\nabla_a \phi\nabla^a\phi-V(x,\phi)\right),
\end{equation}
where $\phi(x)$ is a  scalar field whose behavior is governed by the effective potential $V(x,\phi)$, which must be explicitly dependent on the fixed and nonbackreacting background geometry \eqref{lifmetric}, with coordinates $x_a$, where $a=0,1,\cdots,D$ and whose  metric determinant is $g$. The field equation derived from Eq. \eqref{action} is
\begin{equation}\label{fieldeq}
\Box \phi=\frac{\partial V}{\partial\phi},
\end{equation}
where $\Box=g^{ab}\nabla_a\nabla_b$ is the d'Alembertian operator. In addition, the energy-momentum tensor for the action $S_{(\phi)}$ is expressed as
\begin{equation}\label{emt}
T_{ab}=\nabla_a\phi\nabla_b\phi-\frac{1}{2}g_{ab}\left(\nabla\phi\right)^2-g_{ab}V(x,\phi)
\end{equation}
and, by direct calculation, one can show that it is covariantly conserved -  i.e, $\nabla_{a}T^{ab}=0$. We also have a conserved current given by $J^a=T^{ab}\xi_b$, where $\xi=-\partial_t$ is the timelike  Killing vector arising from the time-translation invariance in Eq. \eqref{timeinv}. The conserved charge related to this current is obtained {\it via}  Stokes' theorem, resulting in 
\begin{equation}\label{consch}
E(\xi)=-\int_\Sigma d^{D-1} x\sqrt{|h|} \eta_a\xi_b T^{ab},
\end{equation}
which is interpreted as the energy of the solution on the background geometry. Here, $\eta_a=-\left(\frac{r}{\ell}\right)^{z}\delta_{a 0}$ is the unit normal vector of the surface $\Sigma$ defined at fixed $t$, with an induced metric denoted by $h_{ij}$ and $h=det\left(h_{ij}\right)$. We must be careful about the convergence of the integral \eqref{consch} by setting conditions on the scalar fields which allow us to obtain finite energy. Furthermore, in analogy to the treatment usually performed when dealing with scalar field solutions in flat spacetime, here we look for minimal energy states (BPS states) such that
\begin{equation}
    E_{BPS}=min\{\left|E(\xi)\right|\},
\end{equation}
which represent ground states of  models derived  from Eq. \eqref{action} on Lifshitz spacetimes for any given potential $V(x,\phi)$.

In order to simplify our approach to the problem and also take advantage of symmetries and the role played by the $r$ coordinate, we restricted our study to static cases where
\begin{equation}
\phi=\phi (r) ~~\text{and}~~V(x,\phi)=V(r,\phi),
\end{equation}
i.e., both the scalar field and the potential depend only radially on the background geometry. By using this choice we significantly simplify the field equations and automatically ensure that the field solutions preserve the set of nonrelativistic symmetries (\ref{nonrelsym}$a$-$c$). The equation of motion for the scalar field \eqref{fieldeq} now becomes
\begin{equation}\label{2ordereq}
\frac{1}{r^{z+D-1}}\left(\frac{r}{\ell}\right)^2\frac{d~}{dr}\left(r^{z+D-1}\frac{d\phi}{dr}\right)=\frac{\partial V}{\partial\phi}
\end{equation}
and, since we are interested in spatially localized solutions, we demand the scalar field to satisfy the boundary conditions
\bes\label{boundcond}\ben
\phi(r\to 0)&=&\phi_{0},~~\phi(r\to\infty)=\phi_{\infty},\\[3pt]
\lim_{r\to 0}\left|\frac{d\phi}{dr}\right|&<&\infty,~~\lim_{r\to \infty}\frac{d\phi}{dr}=0,
\een\ees
where $\phi_0$ and $\phi_{\infty}$ are constants to be determined by the model. Perturbative approaches in the Klein-Gordon equation on Lifshitz spacetimes indicate that stable scalar fields living on Lifshitz backgrounds must satisfy some bounds imposed on the scalar mass from regular boundary conditions  \cite{keeler2014scalar,andrade2013boundary}. However, such issues do not affect the solutions we found here, as our approach is nonperturbative and the scalar fields considered are real. 

\section{BPS formalism}

In this section we present a way to capture BPS solutions in the system discussed so far by building an appropriate first-order formalism from generalizing the approach present in \cite{bazeia2003new} to Lifshitz spacetimes. In order to follow this route, we first note that the integrand of \eqref{consch} is given by
\begin{equation}
    -\eta_a \xi_b T^{ab}=\left(\frac{r}{\ell}\right)^{z}T^0_{~0},
\end{equation}
and that the $00-$component of the energy-momentum tensor \eqref{emt} can be expressed as
\begin{equation} \label{too}
-T^0_{~0}=\frac{1}{2}\left(\frac{r}{\ell}\frac{d\phi}{dr}\mp\sqrt{2V}\right)^{2}\pm \frac{r}{\ell}\frac{d\phi}{dr}\sqrt{2V}.
\end{equation}
Consequently, the scalar potential must be non-negative everywhere on the  geometry and the right-hand side of \eqref{too} has a minimum value when its quadratic term is zero, i.e.,
\begin{equation}\label{1oeq}
\frac{d\phi}{dr}=\pm\frac{\ell}{r}\sqrt{2V}.
\end{equation}
Hence, we can infer that in this model solutions emerge in pairs and also ensure that the weak energy condition,
\begin{equation}\label{wec}
\rho=\left(\frac{r}{\ell}\right)^2\left(\frac{d\phi}{dr}\right)^2=T_{ab}\zeta^a\zeta^b\geq0,
\end{equation}
is satisfied for any unit timelike vector $\zeta^a$. Therefore, the energy density (denoted by $\rho$) of the model remains non-negative everywhere.

Equation \eqref{1oeq} is first order and leads us to minimal-energy solutions, so we can use it as a base to raise up the first-order formalism and achieve analytic BPS solutions.
We need a way to solve Eq. \eqref{1oeq} and with this aim we insert in the model an auxiliary function $W(\phi)$ through the relation
\begin{equation}\label{1ordereq}
\frac{d\phi}{dr}=\pm \left(\frac{\ell}{r}\right)^{\alpha}\frac{dW}{d\phi}, 
\end{equation}
where $\alpha=z+D-1$. In particular, it implies that the scalar potential now becomes
\begin{equation}\label{potmodel}
V(r,\phi)=\frac{1}{2}\left(\frac{\ell}{r}\right)^{2(\alpha-1)}\left(\frac{dW}{d\phi}\right)^2,
\end{equation} 
and one can note that, as expected, it depends on the radial coordinate except for the case with $\alpha=1$ (or, equivalently, $z=2-D$). 

It is easy to show that the solutions of Eq. \eqref{1ordereq} along with the potential \eqref{potmodel} for a given auxiliary function $W(\phi)$ also satisfy the second-order equation \eqref{2ordereq}. In addition, Eq. \eqref{1ordereq} holds scale invariance for $\alpha=1$ and the scalar potential \eqref{potmodel} is invariant under the change $\phi\rightarrow-\phi$. We also have 
\begin{eqnarray}\label{intpot}
\int \frac{d\phi}{W_\phi}=\left\{
\begin{array}{rcl}
\pm\frac{\ell^\alpha}{1-\alpha}\left(r^{1-\alpha}-r_0^{1-\alpha}\right),&\text{if}& \alpha\neq 1,\\[8pt]
\pm\ell\ln \frac{r}{r_0}~~~~~,&\text{if}& \alpha=1,
\end{array}
\right.
\end{eqnarray}
where, for simplicity, we use $W_\phi=dW/d\phi$ and $r_0\geq0$ represents an integration constant. Furthermore, analytical solutions can be found  for the cases where the integral in the left-hand side of \eqref{intpot} can be solved, and if the integration constant is zero the invariant scaling branch of the solutions is wrecked, which suggests that $r_0$ is a preferred point of the system, as well as $r=0$. 

The regularity of these solutions at the boundaries $r=0$ and $r\to\infty$ depends on the values of $\alpha$ and the zeros of the scalar potential. Since we assume that the scalar field and its first derivative are finite on these limits, the way the first-order equation \eqref{1ordereq} is expressed indicates that there may be divergences on the boundary conditions \eqref{boundcond} depending on the values of $\alpha$ if the auxiliary function is not chosen carefully. In particular, to avoid this problem we demand that the auxiliary function $W(\phi)$ to be chosen in such a way that it inserts in the scalar potential $V(r,\phi)$ some set of degenerate minima, which ensures the existence of scalar sectors where nonzero finite-energy solutions can be located.

Under these considerations, a direct calculation shows that the resulting BPS energy of the model becomes
\begin{equation}\label{ebps}
E_{BPS}=\left|\Delta W\right|\omega_{D-2},
\end{equation}
where  $\Delta W= W(\phi_{\infty})-W(\phi_{0})$ and $\omega_{D-2}=\int d^{D-2}x$ is the Euclidean volume related with the $x^i$-coordinates of $\Sigma$. The convergence of $E_{BPS}$ is related to the topological profile of the transverse surface defined by the coordinates $x^i$, which is related to the boundary of $\Sigma$ for fixed $r$, denoted by $\partial\Sigma$. In the standard approach that surface is treated as noncompact; hence $E_{BPS}$ is divergent and we must understand $\left|\Delta W\right|$ as an energy-density term (or a tension) of the field on the background geometry. However, one can also implement a coordinate identification $x^i\to x^i+l^i$ to turn the transverse surface into a compact one (as occurs in \cite{bordo2020geometric,ayon2009lifshitz,ayon2010analytic,deveciouglu2011thermodynamics}, for example) and catch a finite BPS energy which can be interpreted as the total energy of the scalar field on the Lifshitz space. The topological current usually associated with the scalar field, $j^a=\epsilon^{ab}\partial_b\phi$, keeps the same expression that it presents in flat spacetime, but in the model we study here - given the fact that we are dealing with nonasymptotically flat backgrounds - it lead us to divergent quantities. An alternative way to link the asymptotic values of the scalar field to a conserved charge arises by defining another current through an effective vector field stated as the one-form
\begin{equation}
\tilde{A}=\tilde{A}_a dx^a=-\phi(r)\left(\frac{\ell}{r}\right)^{\gamma+1}dt,
\end{equation}
where, for later convenience, we use $\gamma=D-z-3$\footnote{Note that $\alpha+\gamma=2(D-2)$ and $\alpha-\gamma=2(z-1)$, so in 1+1 dimensions we have $\alpha=-\gamma$ and in the $AdS_D$ setup we must have $\alpha=\gamma$.}. By using the above defined field, we can write an antisymmetric tensor $\tilde{f}^{ab}=\partial_a \tilde{A}_b-\partial_b \tilde{A}_a$ and relate it with the current
\begin{equation}
\tilde{J}^{a}=\nabla_b \tilde{f}^{ab},
\end{equation}
which satisfies $\nabla_a \tilde{J}^a=0=\partial_a \left(\sqrt{-g}\tilde{J}^a\right)$
and has an associated charge given by
\begin{equation}
\tilde{Q}=-\oint_{\partial\Sigma_\infty}\!\!\!\!\! d^{\small{D-2}}x \sqrt{|h^{(2)}|} \eta_a s_b \tilde{f}^{ab}=\frac{(\gamma+1)\phi_\infty}{\ell}\omega_{D-2},
\end{equation}
where $s^a=\left(\frac{r}{\ell}\right)\delta^a_1$ is a unit spacelike nomal vector to the boundary of $\Sigma$ at fixed $r$, denoted by $\partial\Sigma$ (in particular, denoted by $\partial\Sigma_\infty$ at spatial infinity) and with induced metric $h^{(2)}_{ij}=\delta_{ij}\left(\frac{r}{\ell}\right)^{D-2}$, whose determinant is given by $h^{(2)}=det\left(h^{(2)}_{ij}\right)$. We also note that the ratio 
\begin{equation}
    \frac{\tilde{Q}}{E_{BPS}}=\frac{(D-z-2)\phi_\infty}{\ell\left|\Delta W\right|}
\end{equation}
is a finite conserved quantity independent of the topological structure of $\partial\Sigma$.

Lisfhitz spacetimes in general have planar symmetry to hold the set of symmetries \eqref{nonrelsym}. 
However, there are models which leave out a subset of these invariances in order to create horizons with distinct topologies, as occurs in Lifshitz topological black holes (see, for example, \cite{mann2009lifshitz}). The model we study here can also be equipped with different topologies without damaging the BPS formalism or the formation of conserved charges. A direct way to do this is by changing $dx^idx^i\to\hat{\sigma}_{ij}(x^k)dx^idx^j$ in the Lifshitz metric \eqref{lifmetric}, where $\hat{\sigma}_{ij}(x^k)$ denotes the metric of a generic Einstein manifold. One can show that in this case the only significant change in the results obtained so far is the exchange $\omega_{D-2}\to\Omega_{D-2}=\int d^{D-2}x\sqrt{|\hat{\sigma}|}$, with $\hat{\sigma}=\text{det}(\hat{\sigma}_{ij})$, in the conserved charges.

\section{Example: $\phi^4$-model}
\begin{figure}	
	\centering
	\begin{subfigure}
		\centering
\includegraphics[width=7.75cm]{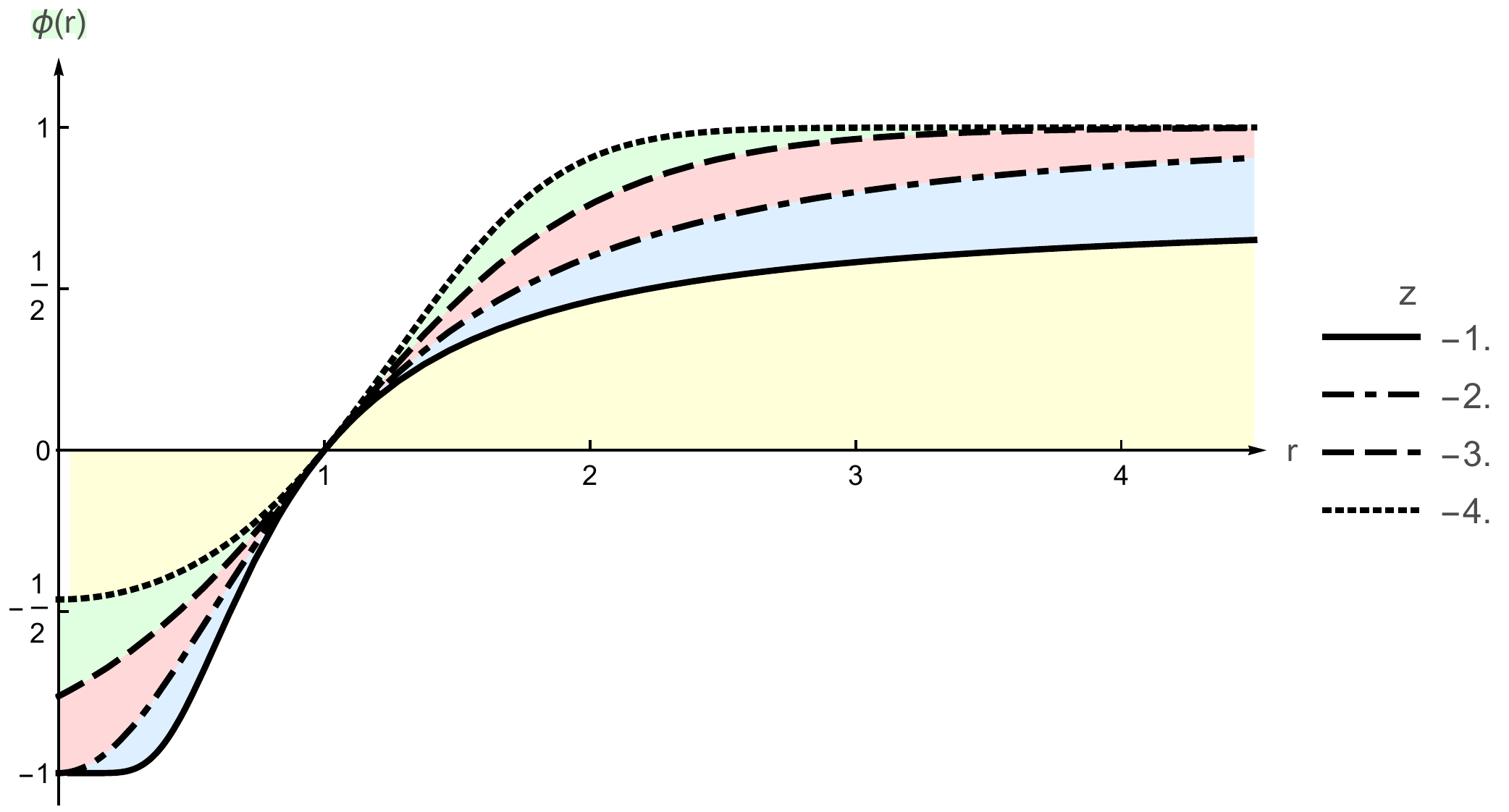}\\
\vspace{2mm}
(a) Scalar field solutions for different values of $z$.\label{fig1a}	
	\end{subfigure}\\
		\vspace{5mm}
	\quad
	\begin{subfigure}
		\centering
\includegraphics[width=7.75cm]{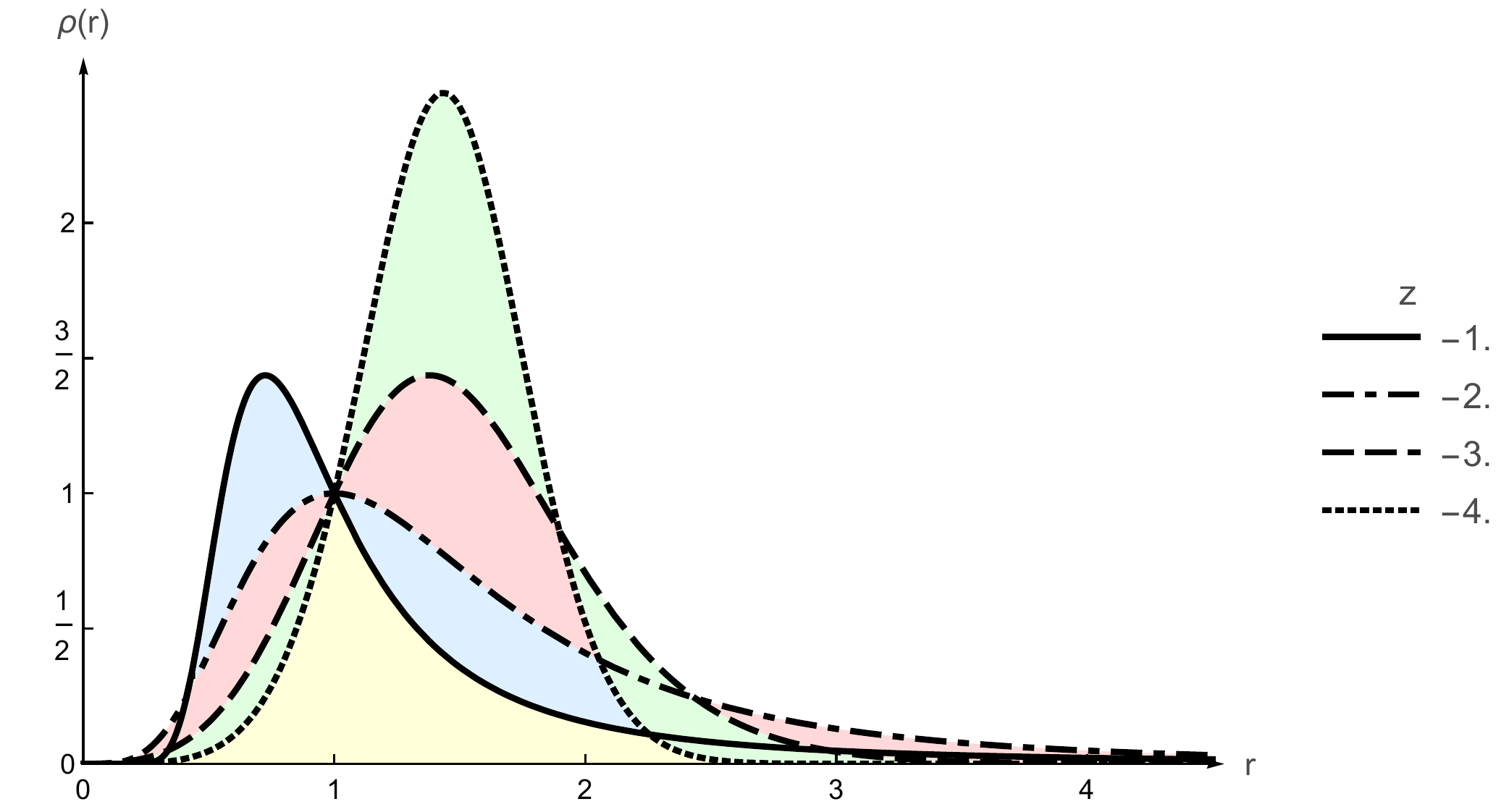}\\
\vspace{2mm}
(b) Energy density for different values of $z$.\label{fig1b}
	\end{subfigure}\\
	\vspace{5mm}
	\quad
	\begin{subfigure}
		\centering
\includegraphics[width=7.75cm]{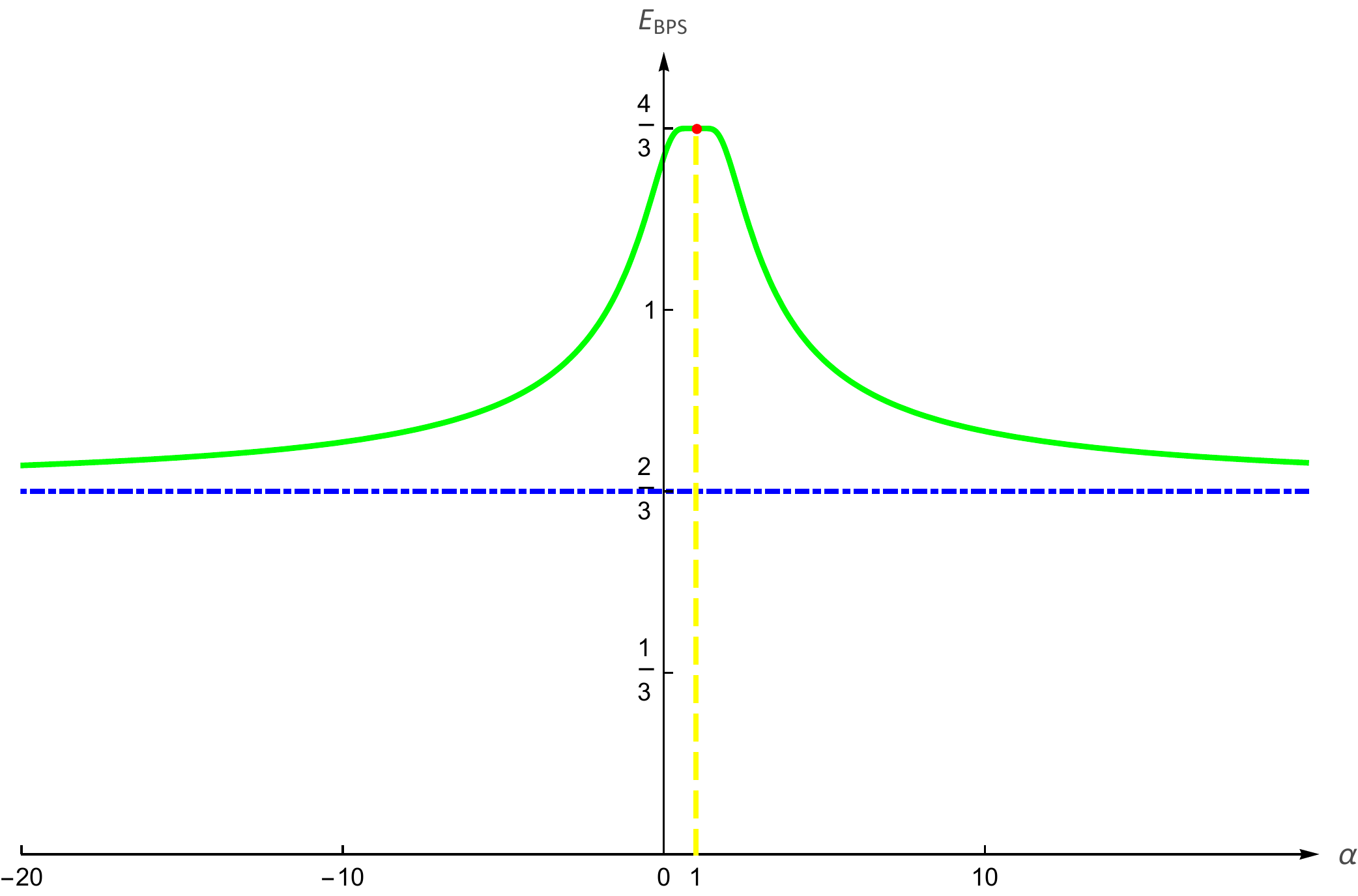}\\
\vspace{2mm}
(c) BPS energy as a function of   $\alpha$.\label{fig1c}
	\end{subfigure}
	\caption{
{\small Spatially-localized profile of the scalar field (a) and the energy density (b) for $D=3+1$ and different values of the dynamical exponent $z$, where we can observe the behavior of the solution for $\alpha>1~ (z=-1)$, $\alpha=1~(z=-2)$ and $\alpha<1~(z=-3~\text{and}~ z=-4)$. The BPS energy as a function of $\alpha$ is depicted in (c).}}
\label{fig1}
\end{figure}
In this section we present an illustrative example of analytical model for scalar field solutions on Lifshitz spacetimes obtained from the formalism introduced and discussed previously. The case under analysis here is the standard $\phi^4$-model, obtained (as usual in the literature) from the auxiliary function $W(\phi)=\phi-\phi^3/3$. The scalar potential for this case  is given by
\begin{equation}
    V(r,\phi)=\frac{1}{2}\left(\frac{\ell}{r}\right)^{2(\alpha-1)}\left(1-\phi^2\right)^2,
\end{equation}
which has a pair of degenerate minima at $\phi=\pm1$. The first-order equation we have to deal with is
\begin{equation}
\frac{d\phi}{dr}=\pm \left(\frac{\ell}{r}\right)^{\alpha}\left(1-\phi^2\right), 
\end{equation}
whose solution becomes 
\begin{eqnarray}
\phi(r)=\left\{\!
\begin{array}{rcl}
\pm\tanh \left(\frac{\ell^\alpha}{1-\alpha}\left(r^{1-\alpha}-r_0^{1-\alpha}\right)\right),&\text{if}& \alpha\neq 1,~~~~~~\\[5pt]
\pm\tanh\left(\ell\ln \frac{r}{r_0}\right)~~~~~~~~,&\text{if}& \alpha=1.
\end{array}
\right.
\end{eqnarray}
One can note that for any value of $\alpha$ we have $\phi(r_0)=0$ and that the scalar field values at the boundaries of the background geometry are  $(\phi_0,\phi_\infty)=(\mp1,\pm\tanh(\tilde{r_0}))$ for $\alpha>1$, $(\phi_0,\phi_\infty)=(\mp1,\pm1)$ for $\alpha=1$ and $r_0\neq0$ and $(\phi_0,\phi_\infty)=(\pm\tanh(\tilde{r_0}),\pm1)$ for $\alpha<1$, with $\tilde{r_0}=\left(\frac{\ell}{r_0}\right)^\alpha\frac{r_0}{|1-\alpha|}$. In addition, for $\alpha\neq1$ the field solutions approach boundaries with exponential behavior (short-range tails), while the  scale-invariant branch of solutions, represented in cases where $\alpha=1$, approach boundaries under power laws (long-range tails). It implies that solutions where $\alpha=1$ approach $\phi_0$ and $\phi_\infty$ with slower rates of change than those that occur with the other cases and, therefore, they have greater energy. The behavior of the scalar field for different values of the dynamical exponent $z$ is illustrated in Fig.\ref{fig1}(a).

The energy density calculated from Eq. \eqref{wec} becomes 
\begin{eqnarray}
\rho(r)=\left\{
\begin{array}{rcl}
\kappa_\alpha \sech^4 \left(\frac{\ell^\alpha}{1-\alpha}\left(r^{1-\alpha}-r_0^{1-\alpha}\right)\right),&\text{if}& \alpha\neq 1,~~~~\\[5pt]
\sech^4\left(\ell\ln \frac{r}{r_0}\right)~~~~~,&\text{if}& \alpha=1,
\end{array}
\right.
\end{eqnarray}
where, for simplicity, we write $\kappa_\alpha=\left(\frac{r}{\ell}\right)^{\!2(1-\alpha)}$. Its behavior is depicted for different values of the dynamical exponent $z$ in Fig.\ref{fig1}(b), where one can observe that the energy density has a spatially-localized profile for any value of $\alpha$ and that for $\alpha=1$ ($z=-2$ for $D=3+1$) it has tails approaching the  boundaries at a slower rate. Moreover, in $r=r_0$ we have $\rho(r_0)=\left(\frac{r_0}{\ell}\right)^{2(1-\alpha)}$, which is an absolute maximum of the energy density for $\alpha=1$. The BPS energy is calculated using Eq. \eqref{ebps} and leads to 
\begin{eqnarray}
\frac{E_{BPS}}{\omega_{D-2}}=\left\{
\begin{array}{rcl}\!\left|\frac{2}{3}+\tanh(\tilde{r_0})-\frac{1}{3}\tanh^3(\tilde{r_0})\right|,\!\!&\text{if}& \alpha\neq 1,~~~~\\[5pt]
4/3~~~~~~~~~~~~~~~~,&\text{if}& \alpha=1.
\end{array}
\right.
\end{eqnarray}
In this case we have $2/3\leq E_{BPS}/\omega_{D-2}\leq4/3$, where
the upper bound is saturated in the case with $\alpha=1$ - as expected - and the lower bound occurs in the limits $|\alpha|\to\infty$ or $r_0\to0$. In Fig.\ref{fig1}(c) we present the evolution of the BPS energy as a function of $\alpha$, where the one can observe that the values of $E_{BPS}$ are symmetrical around the maximum situated at $\alpha =1$. 

\section{Stability}

Finally, let us proceed to study the stability of the model presented so far. In order to perform this task we apply a periodic perturbation around the static scalar field solutions as $\phi(r,t)=\phi(r)+e^{i\omega t}\psi(r)$, which leads us to the stability equation
\begin{equation}
\left(-\Box+\left.\frac{\partial^2 V}{\partial\phi^2}\right|_{\phi=\phi(r)}\right)\psi(r)=\omega^2\left(\frac{\ell}{r}\right)^{2z}\psi(r),
\end{equation}
where $\Box$ represents the D'Alambertian operator in the background geometry. The equation above can be rewritten as an Sturm-Liouville (SL) problem $L\psi=\omega^2\sigma(r)\psi$, where $\sigma(r)=\left(r/l\right)^{\gamma}$ and \begin{equation}\label{eqs_sl}
L=-\frac{d~}{dr}\left(p(r)\frac{d~}{dr}\right)+q(r)
\end{equation}
is an SL operator with
\begin{equation}
p(r)=\left(\frac{r}{l}\right)^{\alpha} ~~\text{and}~~q(r)=\left(\frac{r}{l}\right)^{\alpha-2}\left.\frac{\partial^2 V}{\partial\varphi^2}\right|_{\varphi=\phi(r)}.
\end{equation}
where, as before, $\alpha=z+D-1$ and $\gamma=D-z-3$. In this case (as usually occurs with SL problems), the spectra arising from the stability equations of the models within the systems we study here is constituted by an infinite tower of bound states, but with no zero mode. Each bound energy can be related  to its respective bound state by using the Rayleigh quotient, given by
\begin{equation}
    \omega^2=\frac{\int_0^\infty\left(\frac{r}{\ell}\right)^\alpha\left(\left(\frac{d\psi}{dr}\right)^2+\left(\frac{\ell}{r}\right)^{2}\frac{\partial^2 V}{\partial\phi^2}\psi^2\right)dr}{\int_0^\infty \left(\frac{r}{\ell}\right)^{\gamma}\psi^2 dr}>0.
\end{equation}
It ensures the stability of the solutions, since we do not have eigenstates with negative energy. 

\section{Ending comments} 

In this work we studied the existence of static solitonlike structures in background geometries presenting scaling anisotropy controlled by a dynamical exponent, the so-called Lifshitz spacetimes. Based on the already known procedures for obtaining topological solutions in flat spacetimes, we found stable BPS solutions by setting a first-order formalism from the insertion of an auxiliary function in the model that induces the formation of scalar sectors where we can place the solutions. The results obtained are new, as previous works related to the theme always occur on flat or asymptotically flat spaces and, therefore, the solutions obtained represent deviations from the proposals for generalizing Derrick's theorem for curved spaces presented so far. For a given $D$-dimensional spacetime, the dymanical exponent acts by controlling all aspects of the solution, from the behavior near the boundaries to the total energy of the solution. The scalar solutions that appear in these models approach the frontiers presenting two types of tails depending on the values of the dynamical exponent - which can occur as power laws in configurations where the first-order equations present as scale invariance or as exponential decays for the remaining cases. In this way, the presented results culminate in new systems where one can find classes of well-behaved structures equipped with a first-order formalism suitable for obtaining analytical solutions. 

One can think of several paths for further research based on the study we carried out here. As speculative examples, we can work on inserting fermions or other gauge fields into the model or look for analytical scalar field solutions in other geometries with different scale anisotropies than the one we dealt with above. The fact that conserved quantities depend on the dynamical exponent and that there is a privileged branch of solutions for $\alpha=1$ - which has more symmetries than the others - can provide new features about the behavior of scalar fields on geometries presenting anisotropies and, consequently, also indicate new hints about the properties of its dual operators, given that the approach presented here is completely nonperturbative.

\begin{acknowledgments}
The author would like to thank D. Bazeia for his relevant contributions in improving this manuscript.


\end{acknowledgments}
\subsection*{ERRATUM}
In the stability analysis presented in this paper, I presented the eigenvalues of the stability equation in Eq. (5.1) from the Rayleigh quotient written in Eq. (5.4), from which I concluded the  non-existence of bound states with negative energies - which would ensure the stability of field solutions - and the non-existence of the zero-mode in these systems. This statement contains a mistake, since one can find situations where $\partial^2V/\partial\phi^2<0$ and in these cases it seems that we can have bound states with negative energy. A way to solve this error arises in the study of the factorization of Sturm-Liouville problems \cite{hounkonnou2004factorization}, since the Eq. (5.1) can be factorized as $S^{\dagger}S\psi=\omega^2\psi$, where 
\begin{subequations}
\begin{eqnarray}
S^{\dagger}&=&\left(\frac{r}{\ell}\right)^{z+1}\left(\frac{d~}{dr}+\mathcal{W}(r)+\xi(r)\right)\\
S~&=&\left(\frac{r}{\ell}\right)^{z+1}\left(-\frac{d~}{dr}+\mathcal{W}(r)\right),
\end{eqnarray}    
\end{subequations}
with
\begin{subequations}
\begin{eqnarray}
\mathcal{W}(r)&=&\frac{z+1}{r}+\frac{W_{\phi\phi}}{r^{D-2}}\left(\frac{\ell}{r}\right)^{z+1}\\
\xi(r)&=&-\frac{D-2 (z+2)}{r}.
\end{eqnarray}    
\end{subequations}

A direct calculation reveals that $S^\dagger$ and $S$ are adjoint under the inner product
\begin{equation}\label{inner}
    \langle \psi_m,\psi_n\rangle=\int_{r_o}^{\infty}dr \left(\frac{r}{\ell}\right)^{D-z-3}\Bar{\psi}_m(r)\psi_n(r),
\end{equation}
associated with the Sturm-Liouville problem in Eq. (5.1), which implies that the models discussed in the article do not have bound states with negative energies and therefore we have assured radial stability. Furthermore, one can also calculate the zero-mode  $\left(\psi_0\right)$ from the equation $S\psi_0=0$, given by 
 \begin{equation}
    \psi_0=c\exp{\left(\int dr \frac{W_{\phi\phi}}{r^{D-2}}\left(\frac{\ell}{r}\right)^{z+1}\right)},
\end{equation}
where $c$ denotes a normalization constant. This mistake was noted by F. A. Brito while discussing the extensions of the article's central idea to other background geometries in \cite{moreira2022localized}. 

\bibliography{biblio}

\end{document}